\documentclass[aps,epsfig,prl, twocolumn, superscriptaddress,showpacs,
showkeys,email]{revtex4}
\usepackage{graphics}
\begin{document} 

\title{\bf Folding of small proteins: A matter of geometry?}
\author{PFN Fa\'\i sca}
\email{patnev@alf1.cii.fc.ul.pt}
\author{MM Telo da Gama}
\affiliation{Centro de Fisica Te\'orica e
Computacional da Universidade de Lisboa, Av. Prof. Gama Pinto 2, 1649-003
Lisboa Codex, Portugal}

\begin{abstract}
\bf{We review some of our recent results obtained within the scope of 
simple lattice models and Monte Carlo simulations that illustrate the role 
of native geometry in the folding kinetics of two state folders.} 
\end{abstract}

\pacs{\bf{87.15.Cc; 91.45.Ty}}
\keywords{\bf{kinetics, funnels, contact order, cooperativity, long-range 
contacts}}

\maketitle

\affiliation{Centro de Fisica Te\'orica e
Computacional da Universidade de Lisboa, Av. Prof. Gama Pinto 2, 1649-003
Lisboa Codex, Portugal}

\section{I. Introduction}

During the 1960s the Nobel Laureate Christian Anfinsen showed through 
\textit{in vitro} experiments that globular proteins are capable of
spontaneously self-assemble into their complex, three-dimensional native
(i.e., biologically active) structures through the process of protein
folding~\cite{ANFINSEN}. This finding suggested that the only information
required for a given protein sequence to fold into its unique native state
is the sequence itself, an idea that gave rise to the so-called ``protein
folding'' problem: The prediction of the native fold from the knowledge of
the protein sequence. This question which, in more fundamental terms,
concerns the understanding of the physical principles involved in the
mechanisms of folding and how the latter, together with the observed folding
rates, are encoded in the protein's amino acid sequence has dominated the
folding literature up to the late 1990s.

In the late 1960s Cyrus Levinthal raised a problem related to Anfinsen's
observation that the native state is the global minimimum of the free energy~
\cite{LEVINTHAL}. By using a simple counting argument, Levinthal quantified
the magnitude of the folding process and concluded that proteins would not
fold in a reasonable amount of time if the search for the native
conformation is performed randomly as may be inferred from Anfinsen's
observation, an idea that became known as the Levinthal paradox.

An important contribution to protein folding research, that proposes a new
scenario for folding, is that of the energy landscape theory (ELT), put
forward  by Bryngelson and Wolynes in the late 1980s~\cite{WOLYNES0}. These
authors recognised that akin to spin glasses proteins exhibit frustration.
Indeed, two kinds of frustration may be distinguished in proteins: energetic
frustration, owing to unfavourable interactions present in the native state,
and geometric frustration that leads to energy barriers between structurally
related conformations and results from geometric constraints (e.g., chain
connectivity). A possible scenario is an energy landscape (i.e., the free
energy as a function of one or more conformation coordinates) similar to
that of random heteropolymers (RHP), with many local minima separated by
energy barriers and where high energy barriers are likely to produce kinetic
traps, that is, long-lived low free energy conformations. However,
analytical studies of RHPs together with Monte Carlo simulations provided
evidence that, by contrast with typical RHPs, the energy landscapes of
kinetically foldable lattice proteins are smooth, with a large free energy
gap separating the native state from misfolded conformations~\cite{SALI},
and its overall slope is such that the protein is easily driven down to the
native state (i.e., it must be funnel shaped)~\cite{WOLYNES1}. Lattice
proteins are models that reduce the protein backbone to a string of single
site beads. While it is true that such minimalist models do not capture the
full complexity of real proteins they are non-trivial models that describe
some potentially relevant aspects of protein folding kinetics. See~\cite
{KAYAMIT} for a recent review on computational methods in protein folding
that discusses the strengths and limitations of lattice models.

In addition to providing deep new insights into the protein folding process
the `energy landscape perspective' has stimulated an invaluable synergy
between theoretical and experimental research in the field of protein
folding. In particular, it allowed for ``new interpretations of existing
experiments and has led to the design of new strategies to probe the details
of the folding process''~\cite{DOBSON}. A clear example of successful
theoretical prediction that shaped modern thinking about protein folding
kinetics is that of the concept of nucleation-growth (or -condensation)
mechanism. The latter was firstly discovered by Shakhnovich and his
collaborators in the context of a lattice model and Monte Carlo folding
simulations~\cite{BIOCHEMISTRY} and confirmed by Fersht~\cite{FERSHT} using
a protein engeneering method termed $\phi $-value analysis~\cite{FERSHT2}.
Furthermore, the above mentioned criteria for kinetically foldable
lattice-polymers have been tested for real proteins. The experimental
results suggest that, in spite of their success in predicting biologically
relevant time frames for folding, smooth energy landscapes and large energy
gaps do not account for the remarkable six order of magnitude range
characteristic of the folding rates of real two-state folders~\cite{GILLES}.
By contrast, a parameter of native geometry, that measures the average
sequence separation of contacting residue pairs in the native fold, named
contact order (CO), appears to be strongly coupled to the kinetics of many
small (with $\approx $ up to 120 amino acids) protein molecules exhibiting
two-state folding kinetics. Indeed, Plaxco \textit{et al.}~\cite
{PLAXCO1,PLAXCO2} found a strong correlation ($r=0.92$) between the CO and
the folding rates of 24 single-domain, two-state folders. More recently,
other measures of the native geometry were proposed and were found to
correlate as well as the CO with real two-state folding kinetics~\cite{LRO,
ZHOU}, supporting the idea that the physics underlying the folding mechanism
of these folders, although `encoded' in the protein's primary sequence, may
not depend strongly on its finer details. These observations triggered a
renewed interest in the effects of the native state's geometrical properties
on protein folding, an issue originally addressed by G\={o} and Taketomi in
a pioneering study based on a two-dimensional protein lattice model where
the relative role of local and non-local contacts in the energetics and
kinetics of folding~\cite{GO} were investigated.

In this paper we review some recent results obtained within the scope of
lattice models and Monte Carlo simulations that address the role of native
geometry in determining the folding kinetics of small lattice proteins.

\section{II. Lattice models and Monte Carlo simulations}

\begin{figure}[tbp]
\centerline{\rotatebox{270}{\resizebox{9cm}{12cm}
{\includegraphics{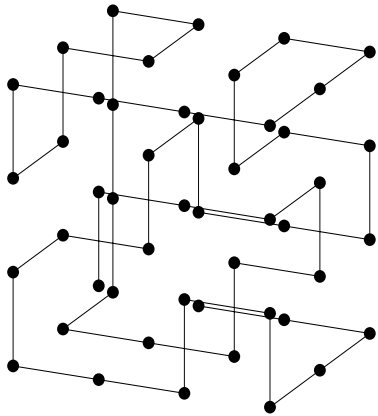}}}}
\caption{A maximally compact conformation (MCC), representing the native
structure, for a 48 bead long chain in a cubic lattice. MCCs, with a maximal
number of contacts between amino acids for each chain length, mimic the high
degree of compactness that characterises real protein native structures.}
\label{figure:no1}
\end{figure}

The lattice-polymer model discretizes space by embedding the protein in a
regular three-dimensional lattice. The protein is reduced to its backbone
structure: amino acids are represented by beads of uniform size, occupying
the lattice vertices and the peptide bond, that covalently connects amino
acids along the polypeptide chain, is represented by sticks, with uniform
length, corresponding to the lattice spacing ( Figure~\ref{figure:no1}). In
order to satisfy the excluded volume constraint only one bead is allowed per
lattice site.

\subsection{The G\={o} model}

In the G\={o} model the energy of a conformation, defined by the set of bead
coordinates $\lbrace \vec{r_{i}} \rbrace$, is given by the contact
Hamiltonian 
\begin{equation}
H(\lbrace \vec{r_{i}} \rbrace)=\sum_{i>j}^N \epsilon \Delta(\vec{r_{i}}-\vec{%
r_{j}}),  \label{eq:no1}
\end{equation}
where the contact function $\Delta (\vec{r_{i}}-\vec{r_{j}})$, is unity only
if beads $i$ and $j$ form a non-covalent native contact, i.e., a contact
between a pair of beads that is present in the native structure, and is zero
otherwise. The G\={o} potential is based on the idea that the native fold is
very well optimised energetically. Accordingly, it ascribes equal
stabilizing energies ($\epsilon <0 $) to all the native contacts and neutral
energies ($\epsilon =0$) to all non-native contacts.

\subsection{The Shakhnovich model}

By contrast with the G\={o} model, which ignores the protein's chemical
composition, the Shakhnovich model~\cite{SHAKHNOVICH} addresses the
dependence of protein folding dynamics on the amino acid sequence by
considering interactions between the $20$ different amino acids used by
Nature in the synthesis of real proteins. Accordingly, the contact
Hamiltonian that defines the energy of each conformation is such that 
\begin{equation}
H(\lbrace \sigma_{i} \rbrace,\lbrace \vec{r_{i}} \rbrace)=\sum_{i>j}^N
\epsilon(\sigma_{i},\sigma_{j})\Delta(\vec{r_{i}}-\vec{r_{j}}),
\label{eq:no2}
\end{equation}
\noindent where $\lbrace \sigma_{i} \rbrace$ represents an amino acid
sequence, and $\sigma_{i}$ stands for the chemical identity of bead $i$. In
this case both native and non-native contacts contribute energetically to
the folding process. Therefore, the contact function $\Delta$ is $1$ if any
two beads $i$ and $j$ are in contact but not covalently linked and is $0$
otherwise. The interaction parameters $\epsilon$ are taken from the $20
\times 20$ Miyazawa-Jernigan (MJ), derived from the distribution of contacts
of real native proteins~\cite{MJ}.

\subsection{Simulation details}

To mimic protein motion in simple lattice models, a Monte Carlo (MC)
algorithm is used together with the kink-jump move set. This means that
local random displacements of one or two beads are repeatedly accepted or
rejected in accordance with the Metropolis rule~\cite{METROPOLIS}. A MC run
starts from a randomly generated unfolded conformation (typically with very
few native contacts) and the folding dynamics is traced by following the
evolution of the fraction of native contacts, $Q = q/Q_{max}$, where $Q_{max}
$ is the total number of native contacts for each chain length, and $q$ is
the number of native contacts at each MC step. Kinetic quantities such as
the folding time, $t$, is taken as the first passage time (FPT), that is,
the number of MC steps that corresponds to $Q = 1.0$. All native structures
considered are maximally compact cuboids found by homopolymer relaxation~
\cite{PFN2}.

The folding dynamics is studied at the so-called optimal folding
temperature, the temperature that minimizes the folding time as measured by
the mean FPT~\cite{SHAKHPRL, JCPSHAKH, CIEPLAK, PFN1}.

The sequences studied within the context of the Shakhnovich model were
prepared using the design method developed by Shakhnovich and Gutin (SG)~
\cite{SGDM} based on random heteropolymer theory and simulated annealing
techniques.

\section{III. Chain length as a determinant of folding kinetics}

There is empirical evidence that the protein's chain length, $N$, grossly
determines the kinetics of folding according to the following rule: for $N$
up to $\approx 100$ amino acids the kinetics is two-state, while for larger $%
N$ protein chains typically exhibit non-two-state kinetic behaviour~\cite
{OAS} (there are of course exceptions to this rule and it is possible to
find proteins with chain lenght $N \leq 110$ that fold via a three-state
kinetics~\cite{JACKSON}). When the kinetics is two-state, folding proceeds
in the absence of any observable intermediates and there is a single
transition state associated with one major free energy barrier separating
the native from the unfolded conformations~\cite{JACKSON}.

\begin{figure}[tbp]
\centerline{\rotatebox{270}{\resizebox{9cm}{9cm}{%
\includegraphics{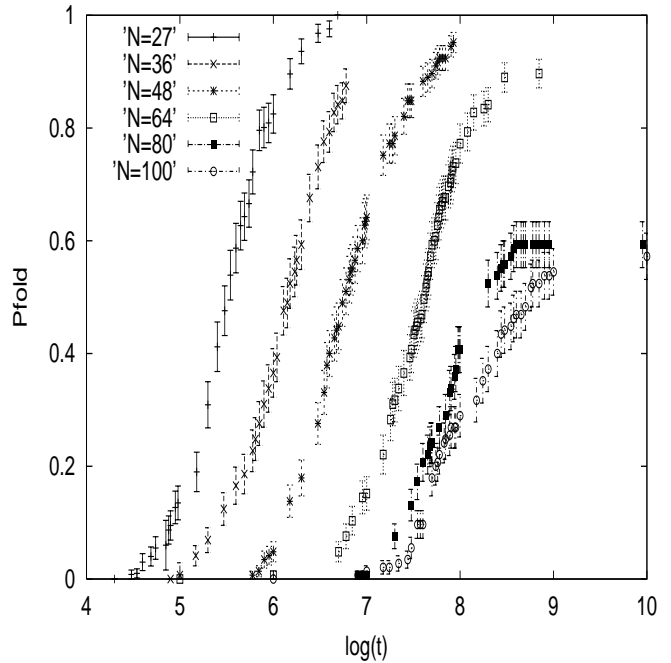}}}}
\caption{Dependence of the folding probability, $P_{fold}$, on $log_{e}(t)$. 
$P_{fold}$ was computed as the fraction of folding simulations that ended up
to time $t$ normalized to the total number of MC runs performed for each
chain length. For $N$ up to 64 the curves are consistent with asymptotic
values of $P_{fold} \rightarrow 1$. For $N \geq 80$ this dynamical behaviour
breaks down and the asymptotic value of $P_{fold}$ decreases considerably~
\protect\cite{PFN1}.}
\label{figure:no2}
\end{figure}

What do lattice models tell us about the role of chain length in the protein
folding kinetics?
Results obtained by one of us~\cite{PFN1} in the context of the Shakhnovich
lattice-polymer model showed that, depending on the chain length, two
dynamical regimes for folding can be identified (Figure~\ref{figure:no2}).
In the regime found for $N \ge 80$ the folding performance depends on the
native state's structure, with certain structures being kinetically more
accessible (i.e., more easily foldable) than others. This observation is in
agreement with results reported in Ref.~\cite{PNASLARGER} where it is found
that, for 125 bead long proteinlike sequences, efficient folding depends on
structural features of the native state (e.g., the distribution and position
of contacts in the native structure). While it is likely that kinetic traps
dominate folding in this regime (akin to what happens in the folding of real
longer proteins) we have not conducted our MC simulations long enough to
conclude that this is actually the case.

\begin{figure}[tbp]
\centerline{\rotatebox{270}{\resizebox{8cm}{9cm}{%
\includegraphics{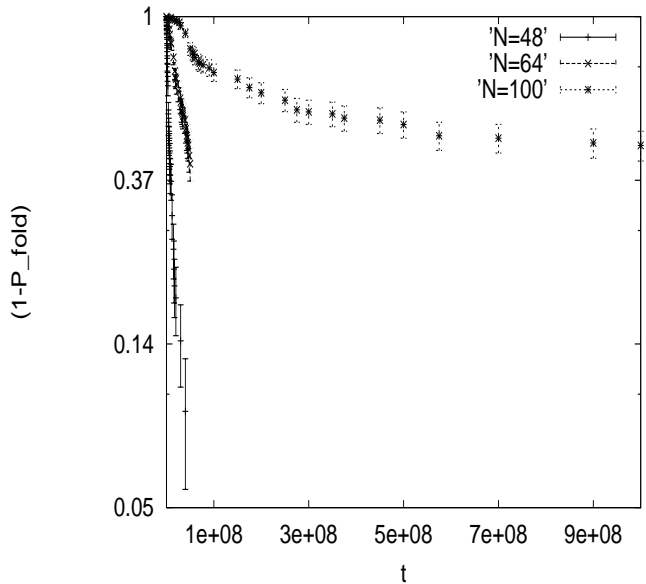}}}}
\caption{Evidence for single exponential kinetic relaxation for $N=48$ and $%
N=64$ but not for $N=100$. The correlation coefficient between the
logarithmic fraction of unfolded chains (i.e., the `reactant concentration')
and the folding time $t$ is $r=0.99$ for $N=48$ and $N=64$~\protect\cite
{PFN1}.}
\label{figure:no3}
\end{figure}

For $N <80$, as in real two-state proteins, we have found that kinetic
relaxation is well described by a single exponential law (a distinguishing
feature of a two-state kinetic process) with the reactant concentration (the
equivalent in our simulations to the fraction of unfolded chains) being
proportional to $\exp^{-k t}$ where $k$ is the so-called relaxation rate
constant (Fig.~\ref{figure:no3}).


As for the dependence of folding time on the chain length we have found a
scaling law of the type $t\sim N^{5}$~\cite{PFN1} while, for the same model,
Gutin \textit{et al.}~\cite{SHAKHPRL} have reported that $t$ and $N$ scale
as $t\sim N^{4}$. For the G\={o} model, on the other hand, a weaker
dependence of $t\sim N^{3}$ has been observed~\cite{SHAKHPRL, JCPSHAKH}.
These findings are in broad agreement with Thirumalai's theoretical
prediction that, near the point of thermodynamic equilibrium between the
unfolded coil and the native fold, i.e., at the folding transition
temperature, the folding time and protein size scale as $t\sim N^{\lambda }$
with $\lambda $ between 3.8 and 4.2~\cite{THIRUMALAI}. For real proteins,
however, it has been shown that the dependence of the folding time on
protein chain length is weak (r = 0.16)~\cite{PLAXCO1}, and recent results
by Galzitskaya \textit{et al}~\cite{GALZITSKAYA} suggest that protein size
is the main determinant of folding rates for multi-state proteins only. We
note that lattice simulations use single or double residue MC move sets
which may differ very significantly from the collective motions of large
portions of the chain that can occur in continuum space. Thus the dependence
of the folding times of  lattice polymers on the chain length could differ
significantly from that observed in real proteins. This shortcoming of the
model may be particularly relevant for  large protein molecules.

\section{IV. Native geometry as a determinant of folding rates}

In 1998 Plaxco \textit{et al.}~\cite{PLAXCO1} proposed the relative contact
order parameter, CO, as a simple empirical measure of the native structure's
geometric complexity. The CO measures the average sequence separation of
contacting residue pairs in the native structure relative to the chain
length of the protein and is defined as 
\begin{equation}
CO=\frac{1}{LN}\sum_{i,j}^N \Delta_{i,j}\vert i-j \vert,  \label{eq:no3}
\end{equation}
where $\Delta_{i,j}=1$ if residues $i$ and $j$ are in contact and is 0
otherwise; $N$ is the total number of contacts and $L$ is the protein chain
length. In a subsequent study Plaxco and co-workers reported a rather strong
correlation ($r = 0.92$) between the CO parameter and the logarithmic
folding rates of 24 single domain, two-state folders~\cite{PLAXCO2},
suggesting that the native's state geometry could be the major determinant
of two-state folding kinetics.

Do lattice models reproduce the experimentally observed correlation between
folding rates and the CO? We have recently addressed this question in the
context of the Shakhnovich model. A set of 20 target native structures,
selected in order to cover the range of CO observed in real proteins, was
investigated for chain lengths $N =36, 48, 54, 64$ and 80. A correlation of $%
r=0.70-0.79$ was found between increasing CO and longer logarithmic folding
times for chain lengths $N \geq 54$~\cite{PFN2}. There is a potential
shortcoming of using the MJ potential to investigate contact order
correlations with folding rates in lattice models. Indeed, it was shown
recently that the energy landscapes of real two-state folders are relatively
smooth~\cite{GILLESPIE, ONUCHIC} while the folding dynamics associated with
the MJ potential is prone to energetic traps (that result from competing
interactions between pairs of beads) that may lead to considerably rough
energy landscapes. Moreover, if the geometry-related kinetics of real
two-state folders is favoured by their smooth energy landscapes and if the
relevant topological effects are subtle they could be masked by stronger
kinetic effects arising in sequence-specific models such as the Shakhnovich~
\cite{JEWETT}. Following this line of reasoning Jewett \textit{et al.}~\cite
{JEWETT} studied the correlation of the folding time on the contact order
parameter in the context of the G\={o} model. Since the G\={o} potential
considers only attractive interactions between native contacts the
corresponding folding dynamics is prone to geometric traps only, i.e., traps
that result from chain connectivity and the geometry of the native fold.
Thus, as real two-state proteins, lattice-polymers modelled by G\={o} and
G\={o}-type interaction schemes exhibit relatively smooth energy landscapes
making them particularly suitable models to investigate the role of the
native state's geometry in the folding kinetics. Notwithstanding, Jewett 
\textit{et al.} found a very poor correlation, $r \approx 0.23$, between
logarithmic folding times and the CO in a pool of targets comprising 97
different native structures with chain length $N=27$. A modified version of
the original G\={o} model, characterized by a non-linear relationship
between the native state's energy and the number of native contacts formed
during folding, i.e., that enhances folding cooperativity, exhibits a
stronger correlation (r=0.75) between the logarithmic folding time and the
contact order~\cite{JEWETT}. Although this correlation is weaker than the
correlation observed experimentally (and lower than that found by us for the
Shakhnovich model) this finding suggests that the physical mechanism
underlying geometry-dependent kinetics may be that of cooperativity.

In a recent study we have identified different folding mechanisms,
distinguished by different cooperativities, \ for the Shakhnovich model~\cite
{PFN3}. For low contact order structures we observed that the building up of
the native fold may occur in a gradual manner, where the amount of native
structure increases in a continuous way as time evolves, or in a more abrupt
(or cooperative) way where a significant portion of native structure emerges
only during the late stages of folding. By contrast,  the folding of
intermediate and high contact order structures is clearly more cooperative,
in the sense described above. Indeed, in the latter case, it is possible to
identify a clear pattern in the formation of the native fold that is driven
by the backbone distance: local contacts (i.e. close in space and in
sequence) form first and long-range contacts (i.e. close in space but
distant in sequence) form progressively later as contact range increases. As
a consequence a monotonic decrease of contact frequency with increasing
contact range is observed and this trend is specific of the folding dynamics
of low contact order structures. These results provide a possible
explanation of the higher correlation found in Ref.~\cite{PFN2} between
contact order and logarithmic folding times for chain length $N\geq 54$ and
predominantly high-CO values. If cooperativity is the essential ingredient
of geometry-dependent kinetics, as the results on modified G\={o} models
suggest, and if the long-range (LR) contacts enhance the cooperativity of
the folding transition it is natural to expect a stronger correlation in
high contact order structures, which have predominantly LR contacts.

\section{V. The role of local and long-range contacts: Revisiting the G\={o}
model}

Contact order is a way to quantify a protein's geometry that accounts for
the average range of amino acid interactions in the native fold. It is
interesting to note that in one of the very first studies that made use of a
lattice-polymer framework to model folding, G\={o} and Taketomi investigated
the role played by local and long-range (LR) inter-residue interactions in
the dynamics of this complex biological process~\cite{GO}. The latter has
actually became one of the most widely-debated issues in the protein folding
literature. As a result of this debate there is now agreement on the fact
that LR interactions play an important role in stabilizing the native fold {
\cite{GO,GUTIN,BAKER1,GROMIHA1}} but there is no consensus on their role in
the folding kinetics. For example, early results of G\={o} and Taketomi {
\cite{GO}} for a 49-residue chain on a two-dimensional square lattice
suggest that local interactions accelerate both the folding and unfolding
transitions. In Ref.{\cite{MOULT}} Unger and Moult have studied optimised
heteropolymer sequences with chain length $N=27$ on a three-dimensional
cubic lattice and concluded that increasing the strength of local
interactions increases the ability of sequences to fold. By contrast,
results obtained by Abkevich \textit{et al.}{\cite{GUTIN}} for the
Shakhnovich lattice-polymer model provided evidence that, under conditions
where the native state is stable, a 36-residue sequence on a
three-dimensional cubic lattice folds to a native structure with mostly LR
contacts two-orders of magnitude faster than a sequence folding to a native
structure with predominantly local contacts. In Ref. {\cite{GOLDSTEIN}}
Govindarajan and Goldstein have used a lattice model in conjunction with
techniques drawn from the theory of spin glasses and found that optimal
conditions for folding are achieved when local interactions contribute
little to the native state's energy.

The finding that the CO is correlated with the folding kinetics of small,
two-state proteins has set a new ground for investigating the role of local
(and LR) contacts in protein folding, one where the effects of native
geometry are taken into account. Moreover, it is natural to address this
issue in the context of the G\={o} model since lattice polymers modelled by
the G\={o} potential exhibit smooth free energy landscapes where the effects
of native geometry dominate. We have revisited the G\={o} model in the light
of these findings to investigate the role of local (and LR) inter-residue
interactions in the dynamics of folding~\cite{PFNTGAN}. We introduced a
parameter $\sigma $ that weights the relative contributions of local and
long-range interactions to the total energy of the native state. When $%
\sigma =0$ all LR native interactions are `switched-off' and only local
interactions contribute to the total energy of a conformation. The opposite
situation is observed when $\sigma =1$. We considered two energy
parametrizations, one at fixed native state's energy, and  three target
structures, with different geometries. We found that, when the native
state's energy varies with $\sigma $, the native fold always exhibits the
highest occupation probability, a measure of its stability, and that, with
the exception of one structure, the latter lies between 0.6 and 0.9~\cite{PFNTGAN}.

\begin{figure}[tbp]
\centerline{\rotatebox{0}{\resizebox{10cm}{10cm}{%
\includegraphics{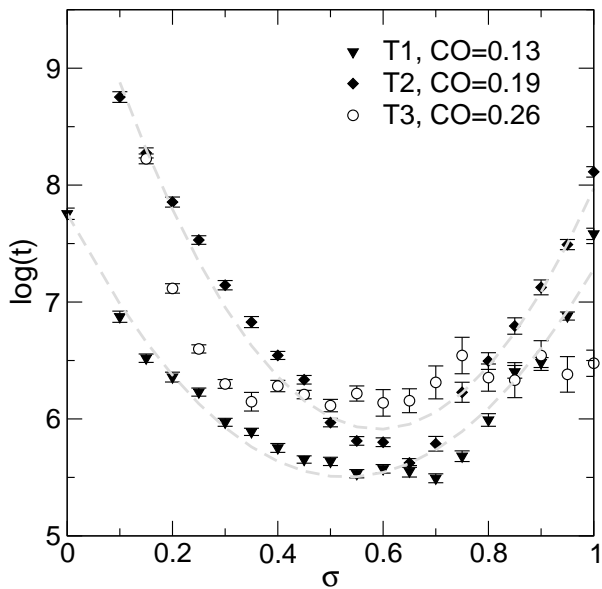}}}}
\caption{Dependence of the logarithmic folding time, $\log_{10}(t)$, on the
long-range interaction parameter, $0 \le \protect\sigma \le 1$ for three
target geometries at fixed native state's energy. The fraction of LR
contacts is 0.77 for target $T3$, while for targets $T1$ and $T2$ it is $0.48
$ and $0.42$, respectively. For the three structures the folding time
increases considerably faster when $\protect\sigma $ decreases than when $%
\protect\sigma $ increases away from the respective minima. The behaviour
observed for $T3$ is trivial and results from its considerably high content
in LR contacts. In the limit of $\protect\sigma=0$, the structure is forced
to fold with only 20 per cent of its native interactions and this results in
folding failure. The results obtained for the low- and intermediate-CO
target structures, $T1$ and $T2$, are more interesting. The corresponding
curves are qualitatively similar but a closer inspection reveals an
important difference, namely: for $\protect\sigma < 0.5$ the dependence of
the folding time on $\protect\sigma$ is much stronger for the
intermediate-CO structure, $T2$. Indeed, in this case one observes a
remarkable three-order of magnitude dispersion of folding times, ranging
from $\log_{10}(t_{min})=5.76 \pm 0.05$ (for $\protect\sigma=0.65$) to $%
\log_{10}(t_{max})=8.75 \pm 0.05$ (for $\protect\sigma=0.10$), by contrast
with $T1$ for which $\log_{10}(t_{min})=5.50 \pm 0.08$ (for $\protect\sigma%
=0.70$) and $\log_{10}(t_{max})=7.69 \pm 0.09$ (for $\protect\sigma=0.00$)~\cite{PFNTGAN}.}
\label{figure:no4}
\end{figure}

Our results show that LR interactions play a major role in determining the
folding kinetics of 48-mer three-dimensional lattice polymers modelled by
the G\={o} potential. Indeed, for three target structures, with different
native geometries, we observed a sharp increase in the folding time when the
relative contribution of the LR interactions to the native state's energy is
decreased. However, the kinetic response to a decrease in the relative
contribution of the LR interactions is strongly dependent on the target
geometry. In fact, we have observed a remarkable three-order of magnitude
span in the folding time of G\={o} polymers folding to one of the target
structures studied (Figure~\ref{figure:no4}).

The existence of different kinetic responses as a function of target
geometry has a mechanistic interpretation. We have found that, for a given
target geometry, a geometric coupling exists between local and LR contacts.
When this is the case, the establishment of LR contacts is forced by the
(previous) formation of local contacts. The absence of this geometric
coupling leads to kinetics that are sensitive to the interaction energy
parameters; in this case establishment of the local contacts is not
sufficient to promote the formation of the LR ones if they are strongly
penalized energetically, resulting in longer folding times.

\section{VI. Proteinlike cooperativity: a new challenge for lattice models}

\begin{figure}[tbp]
\centerline{\rotatebox{0}{\resizebox{8cm}{5cm}{%
\includegraphics{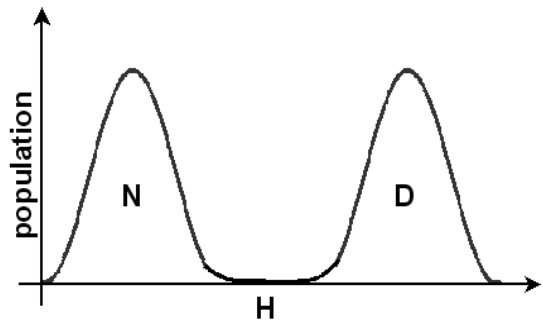}}}}
\caption{Calorimetric criterium: the distribution of enthalpy at the
midpoint of a two-state folding transition is bimodal with very few
molecules having enthalpies between the native (N) and denatured (D) states.}
\label{figure:no5}
\end{figure}

A well-established criterion for two-state thermodynamic cooperativity
observed in folding experiments of real proteins is the calorimetric
criterion introduced in the 1970s by Privalov and co-workers~\cite{PRIVALOV}%
. The calorimetric criterion relates thermodynamic cooperativity to a
bimodal distribution of the enthalpy, describing a conformational population
peaked around the native and the unfolded states and practically zero at
intermediate enthalpies, at the midpoint of the folding transition (Figure~%
\ref{figure:no5}). 
More recently, it was suggested that a kinetic criterion known as the
``chevron-plot" is the definitive hallmark of two-state folding
cooperativity~\cite{ENZYMOLOGY}. Why? The ``chevron-plot" is a V-shaped
graph, that results from plotting the logarithm of $k$, the relaxation rate
constant, as a function of the denaturant concentration (at constant
temperature), where one observes that the folding rate constant, $k_f$,
dominates at low denaturant concentration while the unfolding rate constant, 
$k_u$, dominates at high denaturant concentration. 
\begin{figure}[tbp]
\centerline{\rotatebox{0}{\resizebox{8cm}{6cm}{%
\includegraphics{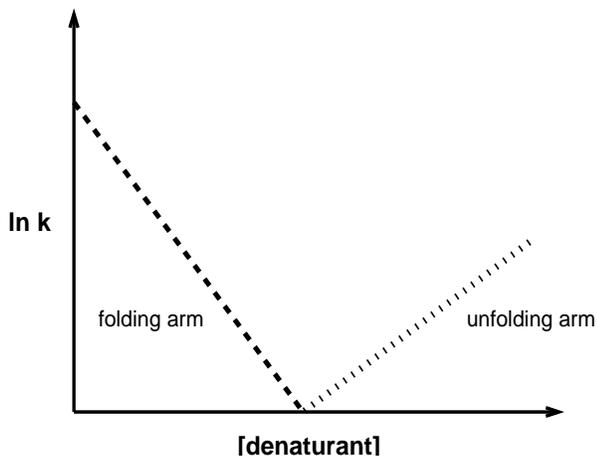}}}}
\caption{A pictorial representation of a `chevron-plot', a distinctive
feature of experimental two-state folding kinetics. The relaxation rate
constant, $k$, is the sum of the folding and unfolding rates constants, $%
k=k_{f}+k_{u}$. The folding rate constant dominates at low denaturant
concentration while the unfolding rate constant dominates at high denaturant
concentration.}
\label{figure:no6}
\end{figure}
The chevron-plot was shown to be a more restrictive criterion for
cooperative folding behaviour than the calorimetric criterion in the sense
that it is not present in all of the two-state proteins that passed the
calorimetric cooperative test. In other words, thermodynamic cooperativity
is a necessary but not a sufficient condition for kinetic cooperativity (see
Ref.~\cite{ENZYMOLOGY} for a recent review on cooperativity principles in
protein folding). The folding of lattice polymers (G\={o} and others)
appears to be relatively non-cooperative in this sense~\cite
{KAYAPRL,ENZYMOLOGY} (note however that, in a broader sense as pointed out
in the previous section, different degrees of cooperative behaviour may be
observed in lattice models.). Based on these observations Kaya and Chan~\cite
{KAYA} suggested that if proteinlike cooperativity is the main drive for
geometry-dependent kinetics, that appears to be lacking in simple lattice
models, it is not surprising that their folding rates are weakly correlated
with contact order. The question is then what happens in these models if
proteinlike cooperativity is enhanced ? Kaya and Chan~\cite{KAYA} addressed
this question in the context of a new G\={o}-type interaction scheme based
on the crucial assumption that a favourable correlation exists between local
conformational preferences and nonlocal interactions, i.e., nonlocal
interactions are stabilised when the chain segments around the native
residues are in their native conformations (the model is suggested by the
experimental observation that secondary structural elements are stable only
in the native fold~\cite{DILL}). This correlation is most simply modelled as
a (non-additive) many-body `attraction' that promotes the folding of large
portions of the native structure which in turn increases the energy gap
between the native and unfolded conformations. Kaya and Chan have found that
this generalized G\={o} model exhibits thermodynamic cooperativity and
linear chevron plots similar to those observed experimentally for real
two-state folders. Moreover the model yields folding rates that are
logarithmically well correlated ($r = 0.94$) with the contact order
parameter. The way in which these many-body correlations arise in general
(or are `encoded' in the primary sequence) remains an open question.

\section{VII. Conclusions}

In the present work we have reviewed some results and concepts that emerged
in the field of protein folding during the last few years, based on results
obtained through Monte Carlo simulations of simple lattice models.

There is experimental evidence that the folding of many small, single
domain, two-state proteins occurs in smooth energy landscapes and that the
folding kinetics is correlated to the geometry of the native state. In this
context G\={o}- and G\={o-}type models are becoming increasingly popular
since lattice-polymers modelled by the G\={o} potential exhibit smooth
energy landscapes where the effects of native geometry dominate. A recent
finding obtained from simulations on these models is that the long-range
contacts play a determinant role in the folding kinetics and that this
effect is geometry-dependent.

However, recent results on the traditional G\={o} potential with additive
pairwise interactions suggest that these models fail to capture the
proteinlike cooperativity of real two-state folding kinetics and the
question of wether realistic folding may be addressed within the scope of
simple lattice models becomes relevant. In particular, the idea that a
geometry-dependent kinetics may result from the type of cooperativity
exhibited by real two-state folders has proven a new challenge for lattice
models. The development of new interaction schemes based on non-additive
many-body interactions leading to G\={o}-type models with protein-like
cooperativity appears to be the next step towards a more realistic lattice
modelling of two-state folding kinetics.

It is an honour and a great pleasure to contribute to this issue in honour
of Ben Widom. Simple lattice models for complex fluids owe a lot to Ben's
imagination and creativity. Thank you Ben for leading the way and Happy Birthday!

\section{VIII. Acknowledgments}

PFNF would like to thank Funda\c c\~ao para a Ci\^encia e a Tecnologia for
financial support through grant number BPD10083/2002.

\end{document}